# Evidence for centrosymmetry breaking and multiferroic properties in the A-site ordered quadruple perovskite $(YMn_3)Mn_4O_{12}$

*M. Verseils[1,†], F. Mezzadri[2,3], D. Delmonte[3], R. Cabassi[3], B. Baptiste[1], Y. Klein[1], G. Calestani[2], F. Bolzoni[3], E. Gilioli[3], and A. Gauzzi[1]*

[1] Institut de Minéralogie, de physique des Matériaux et de Cosmochimie – Sorbonne Université, UPMC, CNRS, IRD, MNHN 4, place jussieu, 75005 Paris, France

[2]Dipartimento di Scienze Chimiche, della Vita e della Sostenibilità Ambientale, Parco Area delle Scienze, 11/a, 43124 Parma (PR), Italy

[3]Istituto dei Materiali per Elettronica e Magnetismo, CNR, Area delle Scienze, 37/a, 43124 Parma (PR), Italy

ABSTRACT By means of single-crystal X-ray diffraction we have revealed the breaking of centrosymmetry when lowering the temperature under $T_s$ = 200 K concomitantly with the setting of a commensurate superstructure in the small A-site quadruple perovskite $YMn_3Mn_4O_{12}$. This results is in agreement with all data already reported for this compounds and solve the previous inconsistency about the Yttrium position. The superstructure is characterized by the appearing of satellite reflections in the single crystal pattern, consistent with an I-centered pseudo-orthorhombic commensurate supercell with $a \approx \sqrt{a_F}$ = 10.4352(7) Å, $b \approx 2b_F$ = 14.6049(9) Å, $c \approx \sqrt{c_F}$ = 10.6961(7) Å and β = 90.110(3)°, where F stands for the "fundamental" high-temperature cell ($a_F \approx c_F \approx$ 7.45 Å, $b_F \approx$ 7.34 Å, and β ≈ 90°). The space-group was unequivocally found to be *Ia*, which is non polar, thus allowing for a non-zero polarization in the material. We then have investigated in detail the pyrocurrent, transport, dielectric and the DC and AC magnetic properties of polycrystalline sample of $YMn_3Mn_4O_{12}$ over a wide temperature range. These measurements clearly highlight several critical temperatures in the material and correlation between the different orders: i) the centrosymmetry is broken at high temperature ($T_S$ = 200K), ii) then the long-range magnetic order of B-sites occurs at $T_{N,B}$ = 108 K and at this same temperature the compound enters in an insulating dielectric state. iii) Finally, a remnant polarization is stabilized concomitantly with a magnetic anomaly at T* = 70 K. We propose that $YMn_3Mn_4O_{12}$ is a peculiar magnetic ferroelectric in which the polar state is driven by short-range magnetic order.

## INTRODUCTION

Manganites with the perovskite structure are one of the most widely studied family of oxides in condensed matter physics. These compounds have attracted a great deal of interest since they host intriguing cross-coupled, lattice, charge, orbital and spin degrees of freedom, leading to very interesting properties. Some of these features, as the colossal magnetoresistance [1-3] and the magnetocaloric effect [4], are already exploited for magnetic sensors and cooling devices while the current research focuses on the possible application of multiferroicity [5-7]. Indeed, this property is very promising for novel concepts of electronic devices, such as non-volatile rapid memories, where magnetization can be switched by an electric field with a modest energy consumption. The ultimate goal is to find materials with both a large polarization and a large magnetoelectric coupling in order to allow the mutual control between the magnetic and ferroelectric orders [8-12]. The quadruple perovskite $(CaMn_3)Mn_40_{12}$, has been reported to hold a large magnetically induced polarization, drawing the attention on this singular perovskite-like structure. This distorted pseudocubic structure, first reported by Marezio et al. [13], hosts two distinct A and A' sites. The driving force of the distortion is the Jahn-Teller effect in the A' site, turning the pristine dodecahedral coordination into an unconventional square planar coordination and driving a very large tilt, up to $\phi = 137°$, of the corner-sharing $MnO_6$ octahedra, much larger than in simple perovskite. This particular angle is responsible for a complex competition of ferromagnetic and antiferromagnetic interactions as predicted by the Goodenough-Kanamori-Anderson (GKA) rules and thus giving rise to the large variety of ground states observed [14]. In a recent work [15], we have reported on the synthesis, crystal and magnetic structure as well as magnetic and thermodynamic properties of a new quadruple perovskite phase, $(YMn_3)Mn_4O_{12}$ (YMnO), where the small yttrium ion is stabilized in the A-site by high pressure synthesis. We found three characteristic temperatures, $T_s = 200$ K, $T_{N,B} = 108$ K and $T^* = 70$ K; $T_{N,B}$ was attributed to the antiferromagnetic ordering of manganese on the B-sites, while $T_s$ was associated to a second-order structural transition described by a significant structural distortion, but no apparent change of symmetry was identified within the resolution of the reported data. No long-range order was identified at T*, characterized by a marked field-dependent anomaly of the DC and AC magnetic susceptibility. Very recently, two articles have been published on YMnO, underlying the large interest of the community in this new phase [16, 17]. [16] reported a new general magnetic structure for rare-earth quadruple perovskite manganites, where both B and A' sites order concomitantly onto a collinear uncompensated ferromagnetic structure when cooling down to the Néel temperature; besides, the occurrence at lower temperature of a second order phase transition is claimed, being related to spin canting at the B sites. However, on one hand the magnetic model supposes a simultaneous long-range ordering of B and A' sites, which is unlikely if the completely different magnetic environment is taken into account. On the other hand, the description of the lower temperature transition requires the mix of three irreducible representations, increasing dramatically the number of the parameters without a significant improvement of the fit of the neutron pattern when compared with the model we already reported [15]. Soon after, Ref. [17] focused onto the structural phase transition occurring at $T_s$, associating it to a displacive transition driven by the underbonding of the Y ions. Still, some inconsistency on the Yttrium position was unsolved. In this article we clarify definitively the nature of the structural transition, reporting for the first time the evidence of a centrosymmetry breaking in YMnO at $T < T_s$., by the setting of a commensurate superstructure, quadrupling the cell volume, clearly identified and characterized by single-crystal XRD data and powder synchrotron diffraction. The non-centrosymmetric nature of the structure is fundamental to understand the possibility of ferroelectricity in the system, suggested by the reported pyrocurrent and dielectric measurements. Finally, magnetic susceptibility and transport measurements are discussed in term of magneto-dielectric properties. The results suggest that YMnO is a peculiar magnetic ferroelectric for which, in spite of the occurring of the centrosymmetry breaking at $T_s = 200$ K, the polarization is stabilized at T* = 70 K by a short-range magnetic order. This behavior, never reported in any $RMn_7O_{12}$, is attributed to the small size of the yttrium ion.

EXPERIMENTAL SECTION

*Sample preparation*: Polycrystalline samples were obtained by a solid state reaction of a stoichiometric mixture of $Y_2O_3$ and $Mn_2O_3$ powders at 9 GPa and 1300 °C for 2 hours using a multi-anvil press, as described in [18]. 3% wt of $YMn_3O_6$ was detected by powder x-rays diffraction as the sole impurity. Single crystals, with a typical size of about 0.05-0.1 mm, were isolated from the polycrystalline samples.

*Single-crystal X-ray diffraction:* Single crystal XRD data were collected at RT and at 150K by using a Bruker D8 Venture instrument, equipped with a Photon CCD area detector, making use of a micro-focused $MoK_\alpha$ radiation. An Oxford cryostream system was used to control temperature below RT. Data reduction, cell refinement, *space*-group determination, scaling and absorption correction [19] were performed using the Bruker APEX3 software [20]. The structure was solved by direct methods using SHELXT [21]. The refinement was then carried out with SHELXL-by full-matrix least squares minimization and difference Fourier methods. All atoms were refined with anisotropic displacement parameters for the RT structure, whereas for the 150K structure the anisotropic refinement was limited to the heavy atoms.

*Pyroelectric current and dielectric constant measurements:* Electrical measurements were carried out on disk obtained by polishing the as-prepared poly-crystalline samples using a *Leica* mechanical polisher. Typical diameter and thickness of the disks were 1 mm and 60-600 μm respectively. Suitable low contact-resistance electrodes were made by covering each disk surface with silver paste. Pyroelectric currents ($I_p$) were measured upon heating up the disk samples using a *Keithley 6517B* electrometer after having poled the sample at a suitable temperature, $T_p$, using a source meter *Keithley 2400*. The capacitance of polycrystalline disks was determined from complex impedance measurement performed with a commercial *HP4824A LCR* meter covering the 20 – $10^6$ Hz frequency range and using an ac voltage of V = 200 mV. For both measurements, the temperature control was achieved by using the liquid helium cryostat of a commercial Quantum Design magnetic properties measurement system (MPMS).

*Magnetic susceptibility:* DC magnetization measurements were carried out in a commercial Quantum Design superconducting quantum interference device magnetometer (SQUID) as a function of temperature and field. AC Magnetization mesurements were performed using a commercial Quantum Design magnetic properties measurement system (MPMS).

*Transport:* Resistance measurements in the 60-300 K range were performed a homemade set-up controlling the temperature by using the liquid helium cryostat of a commercial Quatum Design Physical Property Measurements system (MPMS). The resistivity measurements carried out in the van der Pauw configuration showed low contact resistances, negligible as compared to the sample resistance. This allowed us to perform impedance measurements in the simple two-contact geometry suitable for our disk-shaped samples.

## RESULTS AND DISCUSSION

*Single-cristal and powder X-ray diffraction:*

In our recently published work [15], in spite of the complex dependence of lattice parameters and cell volume on temperature, no symmetry decrease was observed by neutron powder diffraction not only at $T_s$ but also in the whole investigated range, so that the data were refined in the centrosymmetric space group *I2/m* in the previously determined pseudo-cubic monoclinic cell (a≈c≈7.45 Å, b≈7.34 Å, and β≈91°) from 300 down to 4K. On the contrary, in the present work, the occurrence of a structural transition at $T_s$, involving a symmetry breaking, was pointed out. This was made possible on one site by the advantages deriving from use of single crystal diffraction and on the other by the use of the Mo $K_\alpha$ radiation, whose wavelength, close to the absorption edge of Y, amplifies by resonant scattering the scattering factor of the Y atoms, whose displacement is at the basis of the structural transition. Several crystals were preliminary tested In search of the "best sample" to perform the experiment, most samples showing a high degree of mosaicist which suggests the occurrence of a phase transition during the preparation process. In fact, in agreement with other $AMn_7O_{12}$ compounds [22-27], the structure is probably cubic in the high-pressure/high- temperature conditions required for the synthesis and the monoclinic distortion occurring on cooling induces microstructural defects. The best crystal, showing the largest size (50x50x40 $\mu m^3$) coupled with minor mosaicity and twinning effects, was selected and use for a data collection at RT and 150 K. The refinement of the RT data, was used to check the crystal quality, satisfactorily confirmed by the crystal data and refinement parameters reported in Table S1. Noteworthy, the refinement of the site occupancy factors converged to the expected values for all the atoms, with the exception of yttrium, for which a deficiency of about 7% was found. This datum was then confirmed for other examined crystals, indicating this as a characteristic of the examined batch and not of a particular crystal. As previously observed in other quadruple perovskite systems [28, 29] the electroneutrality of the samples is probably guaranteed by a partial oxidation of manganese at the B site, leading to a phase stoichiometry $Y_{0.93}Mn^{III}_3(Mn^{III}_{3.79}Mn^{IV}_{0.21})O_{12}$. Full crystallographic data and refinement parameters are reported as Sup Info. The analysis of the bond distances is consistent with a disordered occupancy of the B sites by the $Mn^{4+}$ ions, which is not able to significantly perturb the orbital order (OO), related to Jahn-Teller distortion of the $Mn^{3+}$ ions with $d^4$ electronic configuration, evidenced by the elongation in the *ac* plane of the apical bonds of the $MnO_6$ octahedra. As shown in Figure 1, the OO gives rise to a checkerboard-like arrangement in the *ac* plane of the long apical bonds, which is repeated in the stacking sequence along *b* by the mirror planes perpendicular to it.

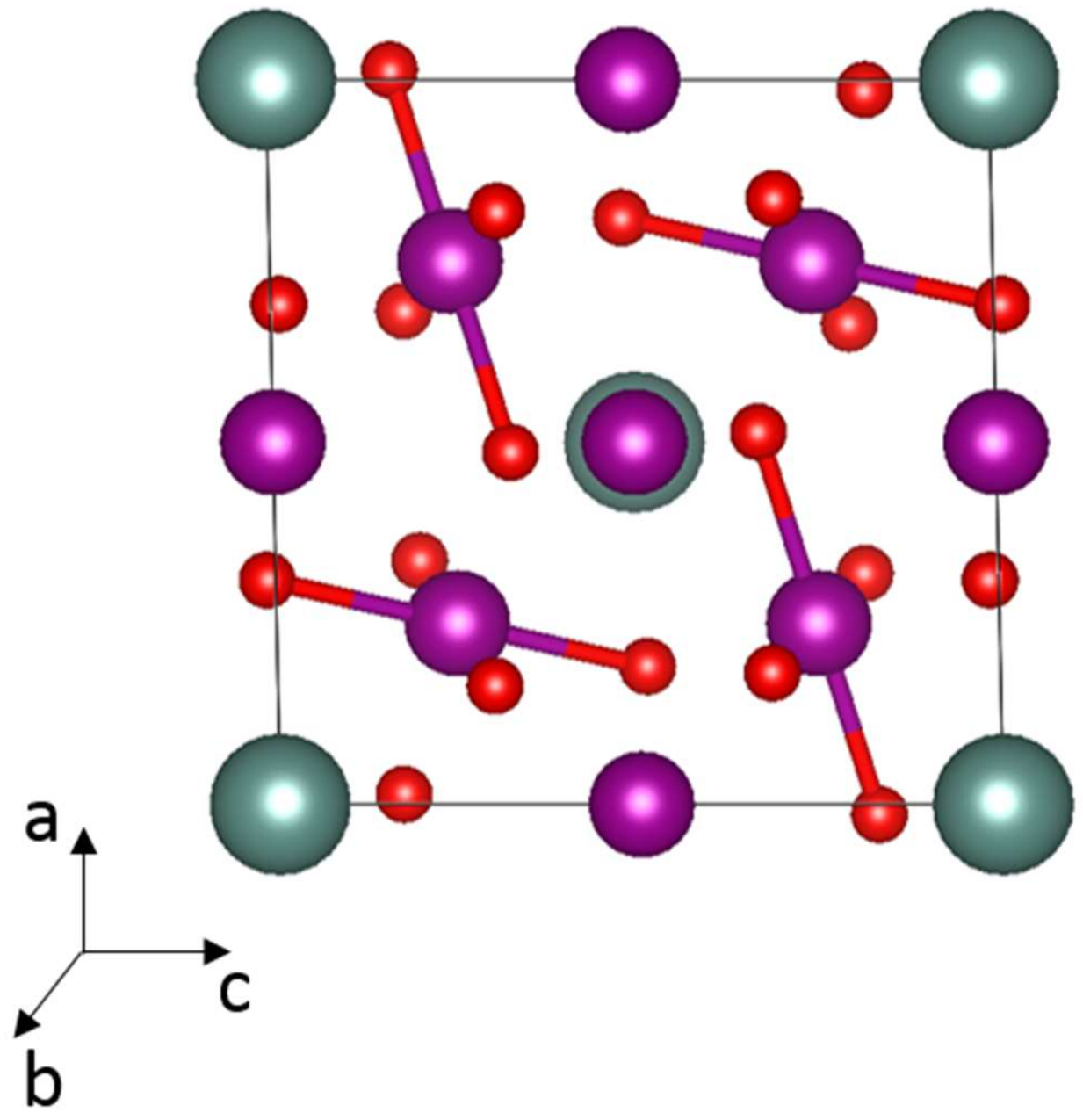

Figure 1 : Projection along b-axis of the $YMn_3Mn_4O_{12}$ cell at RT. Y is represented in dark cyan, Mn in violet, oxygen in red. Only the longer Mn-O bonds are depicted in order to highlight the orbital order (OO).

The behavior of the lattice parameters and crystal volume, previously determined by neutron powder diffraction on cooling from RT to 5K, shows an anomaly at 200 K dramatically modifying the thermal expansion coefficient [15]. However, differently from what observed in powder diffraction, the single crystal pattern is characterized, on cooling below $T_S$ (200 K), by the appearing of satellites reflections (see Supporting Information) consistent with an I-centered pseudo-orthorhombic commensurate supercell with $a \approx \sqrt{}a_F$, $b \approx 2b_F$, $c \approx \sqrt{}c_F$, where F stands for the “fundamental” cell determined at RT. In spite of the fact that unconstrained refinements of the lattice parameters led to small deviations of the unit cell angles from 90°, indicating a possible orthorhombic nature of the structure, a much lower $R_{int}$ was obtained by merging the equivalent reflections in the monoclinic rather than in the orthorhombic Laue class ($R_{int}$ = 0.0217 for *2/m* and $R_{int}$ = 0.116 for *mmm* symmetry) in agreement with systematic absences affecting the *hkl* reflections with *h+k+l=2n+1* and the *h0l* with *h,l=2n+1* that reduces the possible space groups to *Ia* and *I2/a*. The final refinement of the lattice parameters with constrained monoclinic symmetry led to a unit cell with a = 10.4352(7) Å, b = 14.6049(9) Å, c = 10.6961(7) Å, and β = 90.110(3) °. The structure solution, at first worthless attempted in the centrosymmetric space group *I2/a*, was finally found in *Ia* and is represented in Figure 2.

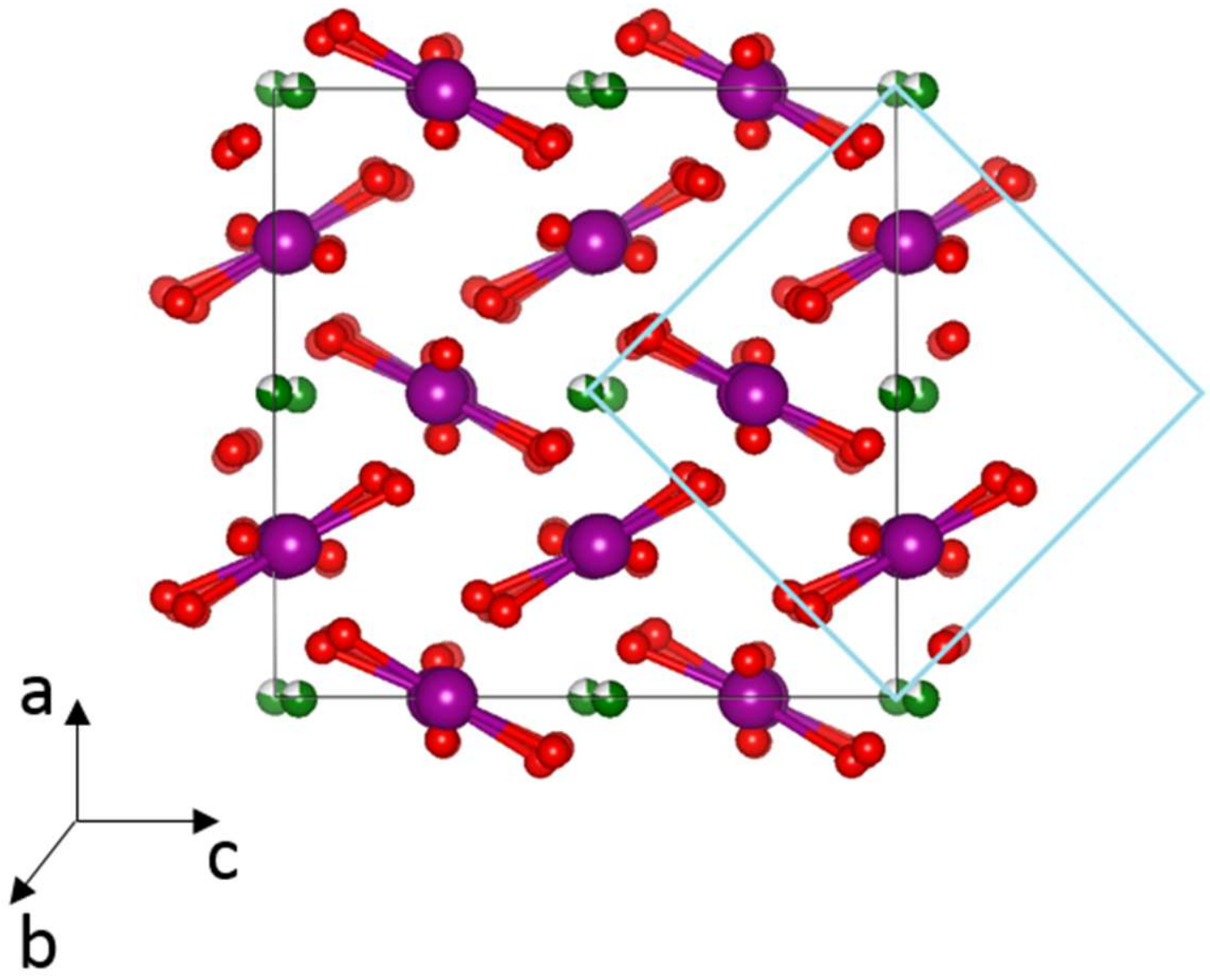

Figure 2 : Projection of the 150K structure along the b axis, where Y, Mn and O atoms are represented by dark cyan, violet and red balls, respectively, and only the longer apical bonds of the octahedral MnO6 units are depicted to illustrate the OO; black and blue lines represent the supercell and the fundamental ones, respectively

To this regard, it should be noticed that the symmetry elements involved in the *I2/a* space group, namely the twofold axes parallel to *b*, are not compatible with the generation of the observed superstructure, which is on the contrary permitted by *Ia*. Very interestingly, one should notice that this space group refers to the point group *m*, which is polar and allows a dipole moment lying in the *ac* plane. Even if the structure is non-centrosymmetric, the deviation of the Mn-O framework from a centric symmetry was found to be scarcely significant, introducing correlations between the refined parameters of Mn an O atoms. In order to reduce the parameters and to limit their correlations, the structure was refined by using anisotropic displacement parameters only for the heavy atoms, whereas the O atoms were refined isotropically. The deviation of the structure from the centric symmetry is given by the displacement of the yttrium atoms from the center of the dodecahedral cage as represented in figure 3, in an ordered sequence that determines the observed superstructure of figure 2. As expected for a ferroelectric sample, the crystal was found to be affected by polar twinning in a 50/50 ratio. However the refinement pointed out, within the single domain, the existence of a partial disorder in the sequence of the yttrium displacements, which has been interpreted as stacking faults of the sequences along the *b* axis or, alternatively, with the existence of polar twinning domains below the coherence length of the used X-ray radiation. The yttrium deficiency observed at RT was detected also in the 150 K data, but in order to simplify the description of the structural disorder, the global occupancy of the Y sites has been constrained, in the final refinement cycles, to those refined in the RT experiment. Crystal data and refinement parameters are reported in Table S1 in Supporting Information. Figure 2 shows a projection along the unique axis *b* of the monoclinic superstructure, where only the longer apical bonds of the $MnO_6$ octahedral units are reported. For the sake of clarity, the cell origin was translated on the Y atoms (in agreement with the restraints imposed by the *Ia* space group) in order to make the comparison with the room temperature fundamental cell as coherent as possible. This figure shows at the same time: a) the preservation of the OO observed at RT; b) the acentric displacements of the Y atoms; c) the almost close overlapping of the Mn-O framework in the different layers; d) the relation between fundamental cell and supercell. Noteworthy, the OO scheme observed at RT is retained and the observed apical elongation can be ascribed to a localization of the charges on the B site. The further change in the thermal expansion coefficient taking place at $T_N$ may be related to magnetostriction taking place at the onset of the magnetic ordering process. The already known increase of the $a_F$ and $c_F$ lattice parameters occurring below $T_S$ and starting at 200 K, is likely mainly ascribed to the shift of the yttrium ions off the center of the coordination cage (figure 2). Within the *ac* plane the Y atoms are displaced along *c* (i.e. along the diagonal of the fundamental cell) in rows alternating along *a*. Simultaneously the Y atoms of adjacent layers are shifted in an antiparallel way along the same direction, doubling the periodicity along *b* with respect to the fundamental cell. The shift of the Y atoms develops also along *b*, but it is fully compensated by the symmetry constraints. Despite the observed displacement scheme may appear at first view as an antiferroelectric pattern, the computation of the electric cell dipole by using a simple point-charge model give a non-zero resultant. The shift of the Y atoms induces an overall slight distortion of the oxygen framework, affecting the regular coordination of the Mn atoms both in octahedral and square planar coordination observed at RT. However, whereas the average Mn-O bond distance remains substantially unaltered for the octahedral A sites, an additional "short" interaction, ranging around 2.4 Å, is observed for half the B sites. The analysis of the crystal structure provides another interesting and unexpected result: by lowering the temperature the cell volume remains practically unchanged in the whole temperature range (0.03% variation between 300 and 50 K). The $YO_{12}$ dodecahedron appears at 50 K strongly distorted, with bonds in the range 2.430(17) - 3.151(18) Å and volume of 45.9630 Å$^3$ while at 300 K bond lengths are more regular: 2.589(3) - 2.731(3) Å and slightly smaller volume (45.2151 Å$^3$).

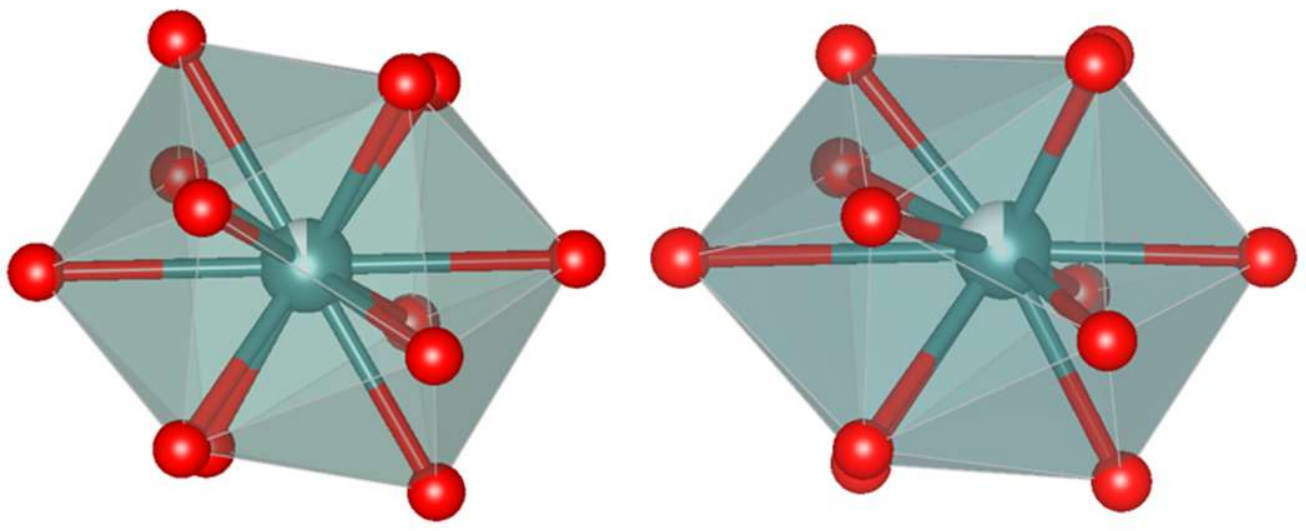

Figure 3 : Representation of the Yttrium dodecahedron cage at RT (left) and at 150K (right). Y and O atoms are represented by dark cyan and red balls respectively.

The low temperature volume increase of the yttrium coordination is likely due to the shift of the cation, as expected for a para-to-ferroelectric transition, in agreement with previous determinations indicating a negative thermal expansion coefficient between about 100 and 200 K, related to an increase of the *a* and *c* cell parameters [15,17]. The present findings suggest that the previous determinations of the magnetic structure of $YMn_7O_{12}$, not taking into account the presence of a superstructure, could be erroneous. On the other hand two different neutron diffraction [15, 16] did not show any superstructure reflection indicating that the Z-dependent atomic scattering factor is crucial to give intensity to the superstructure reflections, i.e. yttrium breaks the symmetry while manganese and oxygen retain more or less the centrosymmetric arrangement observed at RT. Moreover Yttrium has an absorption edge pretty close to the $MoK_\alpha$ energy, thus enhancing its atomic scattering factor due to almost resonant conditions. For this reason it is not surprising that the periodicity of the magnetic ordering, involving the manganese atoms, coincides with the one of the fundamental cell. It should be noticed that if the low temperature diffraction data are refined using the fundamental cell, the correct space group symmetry is *Im* and the yttrium shift must be taken into account as an average phenomenon. However, if magnetic symmetry is taken into account, switching from *I2/m* to *Im* does not modify the number of independent manganese atoms at the B site (2) and the magnetic symmetry is unaffected as the combined effect of *I* centering and *m* plane on a spin sitting on the high symmetry positions *4f* and *4e* is to mimic the presence of the twofold axis. For these reasons the previous neutron diffraction study carried out by Verseils et al. [15] should be considered completely reliable. On the contrary, based on the present structural determination, the space group symmetry proposed by Johnson et al. [16] ($P2_1/n$) is not correct, in particular for what concerns the nuclear crystal structure. On the other hand, the high thermal parameters detected were correctly interpreted as yttrium off-centering, but the phenomenon is not statistical and the presence of a superstructure reveals its periodic character. Conversely, also in this case the number and positions of independent magnetic ions as well as the magnetic symmetry are compatible with those of *Im* and consequently the treatment of the magnetic neutron diffraction should be considered reliable.

*Pyrocurrent and transport measurements:*

To support the above diffraction results revealing a centro-symmetry breaking, we have investigated the electrical properties of YMnO by measuring the pyroelectric current and transport measurements of polycrystalline samples. Figure 4 shows the pyrocurrent peak vs. temperature for a poling electric field of 3.2 x $10^5$ V/m, a warming rate of 1.3K/min and different poling temperatures $T_p$= 48, 55, 75 and 104 K. In all curves we identify the negative contribution of leak current only above $T_{N,B}$ = 108 K. The peak obtained at $T_p$ = 104 and 75 K are comparable, larger and higher than the peak at $T_p$ = 48 and 55 K, indicating that the critical temperature lies between 75 and 55K. Peaks obtained with a high $T_p$ present a shoulder around 45K, suggesting a second critical temperature.

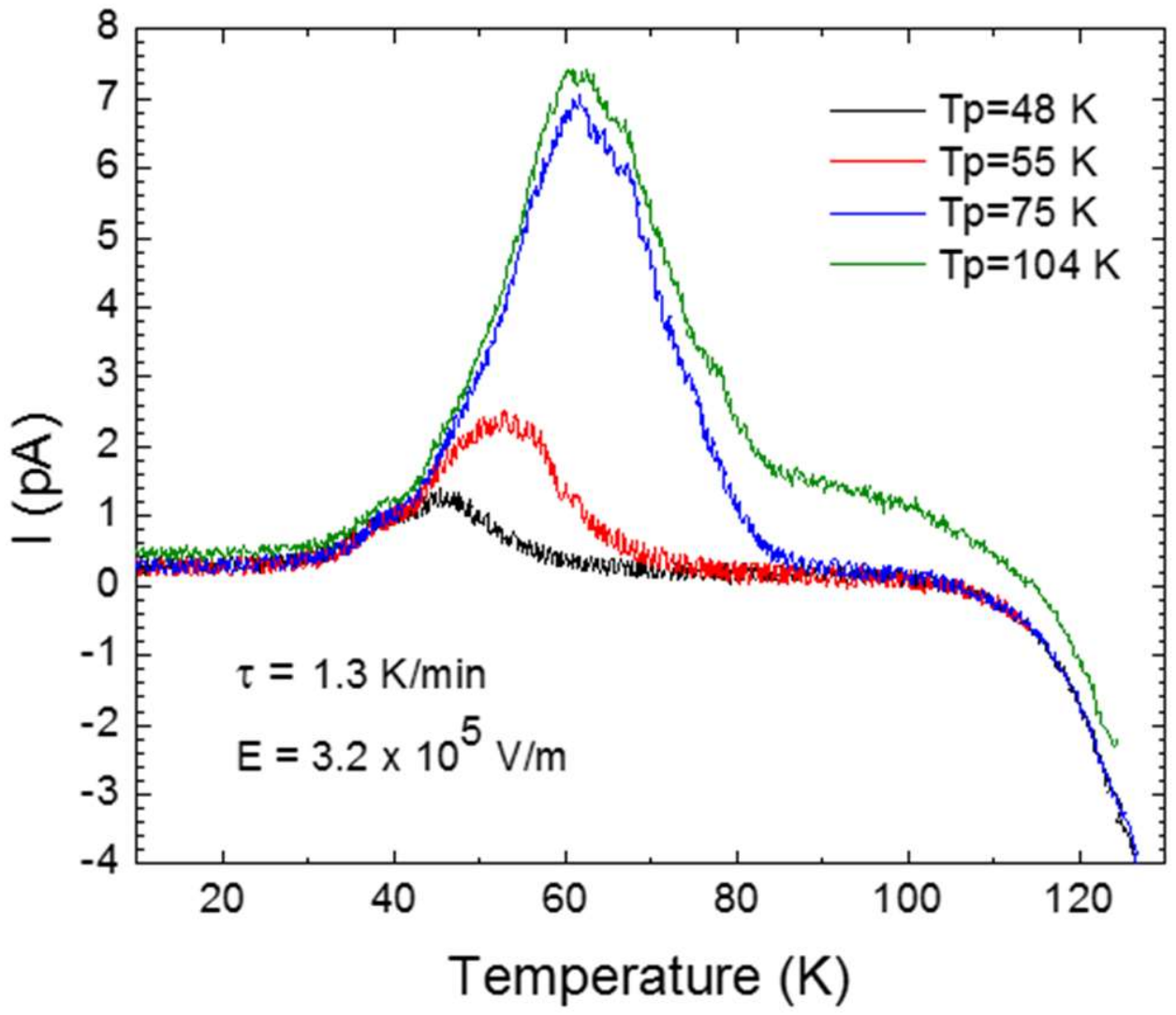

Figure 4 : Pyroelectric current measured on polycrystalline sample of YMnO and obtained for different temperature of poling $T_p$ = 48, 55, 75, 104 K

The pyrocurrent I(T) was also measured under various warming rates on the same sample with constant $T_p$ = 100K and constant poling electric field E = 3.2 x $10^5$ V/m. The polarization curves P, obtained by the integration of the, I(T), curves are shown in Figure 5. The warming rate has almost no influence on the saturation value of P, which consistently falls in the 0.20 – 0.22 μC $cm^{-2}$ interval. In addition, the I and P curves obtained by inverting the sign of poling voltage also change sign without modifying their magnitude, as expected for an intrinsic polarization. As reported in a recent study [30], there is a risk of measuring pyroelectric currents derived from extrinsic effect in polycrystalline samples, such as thermally activated space-charges at grain boundaries. Such extrinsic effects are expected to vary as a function of warming rate, which is not the case in our results as shown in Figure 5; this confirms the intrinsic nature of polarization in YMnO.

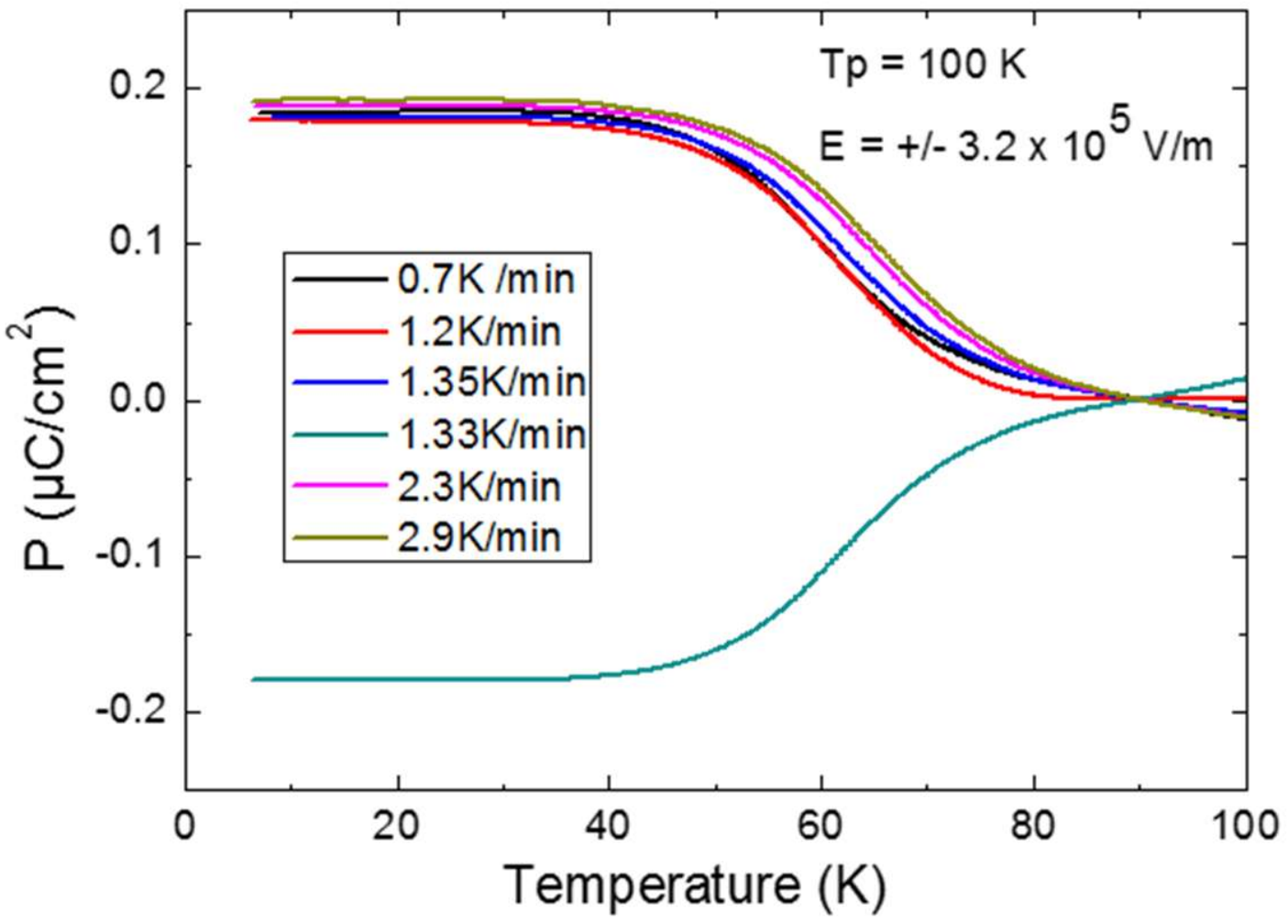

Figure 5 : Polarizations obtained by integration from pyroelectric current measurements performed under various warming rate of 0.7, 1.2, 1.35, 1.33, 2.3 and 2.9 K/min.

Besides, the polarization at 8K reported on figure 6 for several samples with different thicknesses, well describes the full P(E) curve up to the saturation region where the P values level off at $P_{sat}$ = 0.54 μC $cm^{-2}$ for a poling voltage E = 3.5 x $10^5$ V/m. This value is twice larger than the record value previously reported for the spin-spiral magnet $CaMn_7O_{12}$ [31]. However, our attempts to measure the ferroelectric hysteresis loop directly on as grown pellets have been unsuccessful so far and in parallel larger single-crystals are needed for such measurements. We also have performed transport measurements on polycrystalline samples. These measurements can be interpreted in the framework of aforementioned pyrocurrent results.

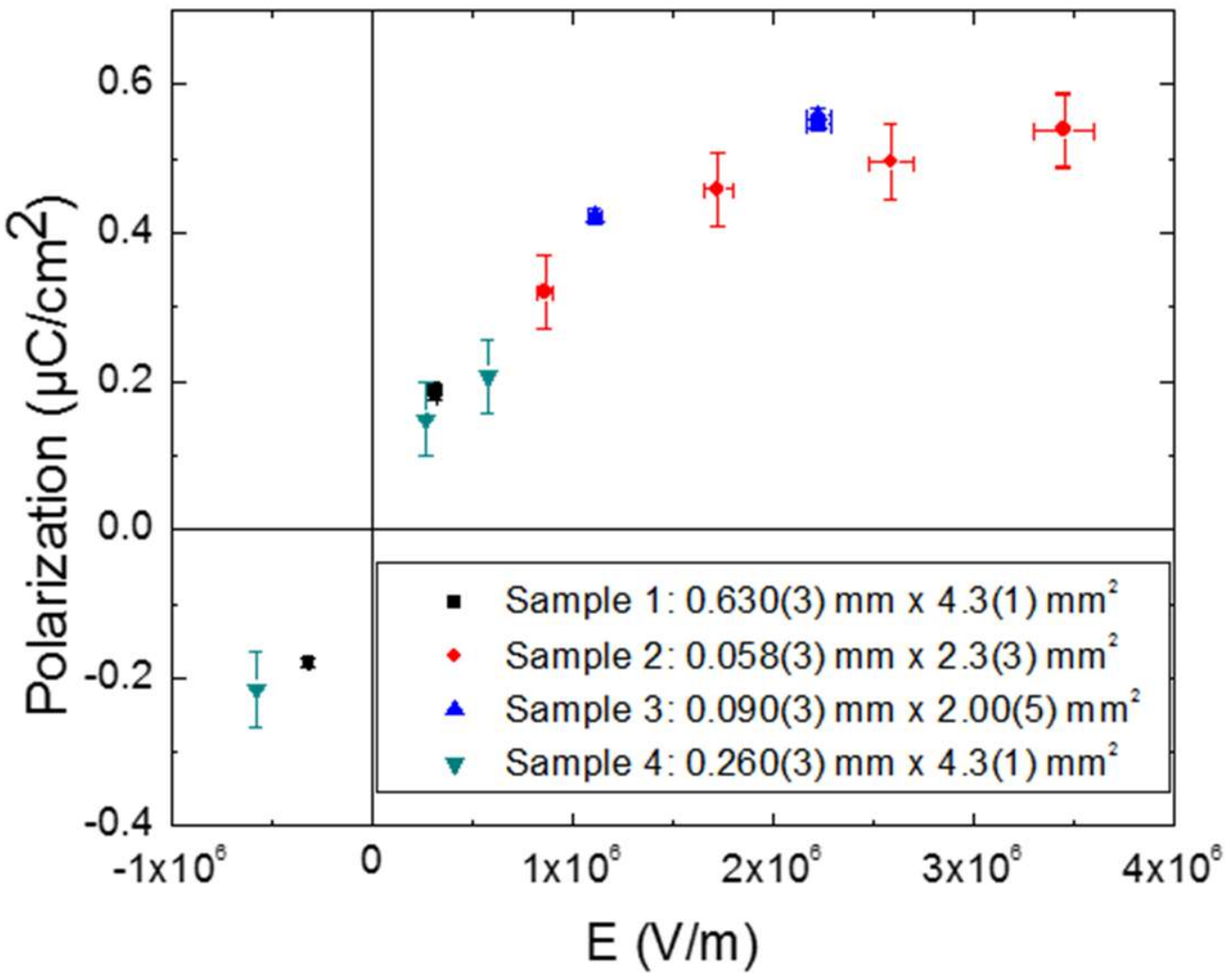

Figure 6 : Low temperature (8K) P-value of polarization obtained as a function of poling electric field E from pyroelectric measurements performed on four YMnO polycrystalline samples under various warming rate with a same poling temperature Tp = 100K.

As shown in Figure 7, the resistivity undergoes two different regimes as a function of temperature: at high temperature (above $T_{N,B}$), the system behaves as a semiconductor with relatively small activation energy ($E_A$=168 meV); below $T_{N,B}$, the system collapses in a dielectric state, progressively losing the linear dependence on the Arrhenius plot as shown in the inset of Figure 7. This behavior is compatible with the pyroelectric measurements which highlights a leak current above $T_{N,B}$, as typically expected for a semiconductor material with significant concentration of carriers in the conduction band.

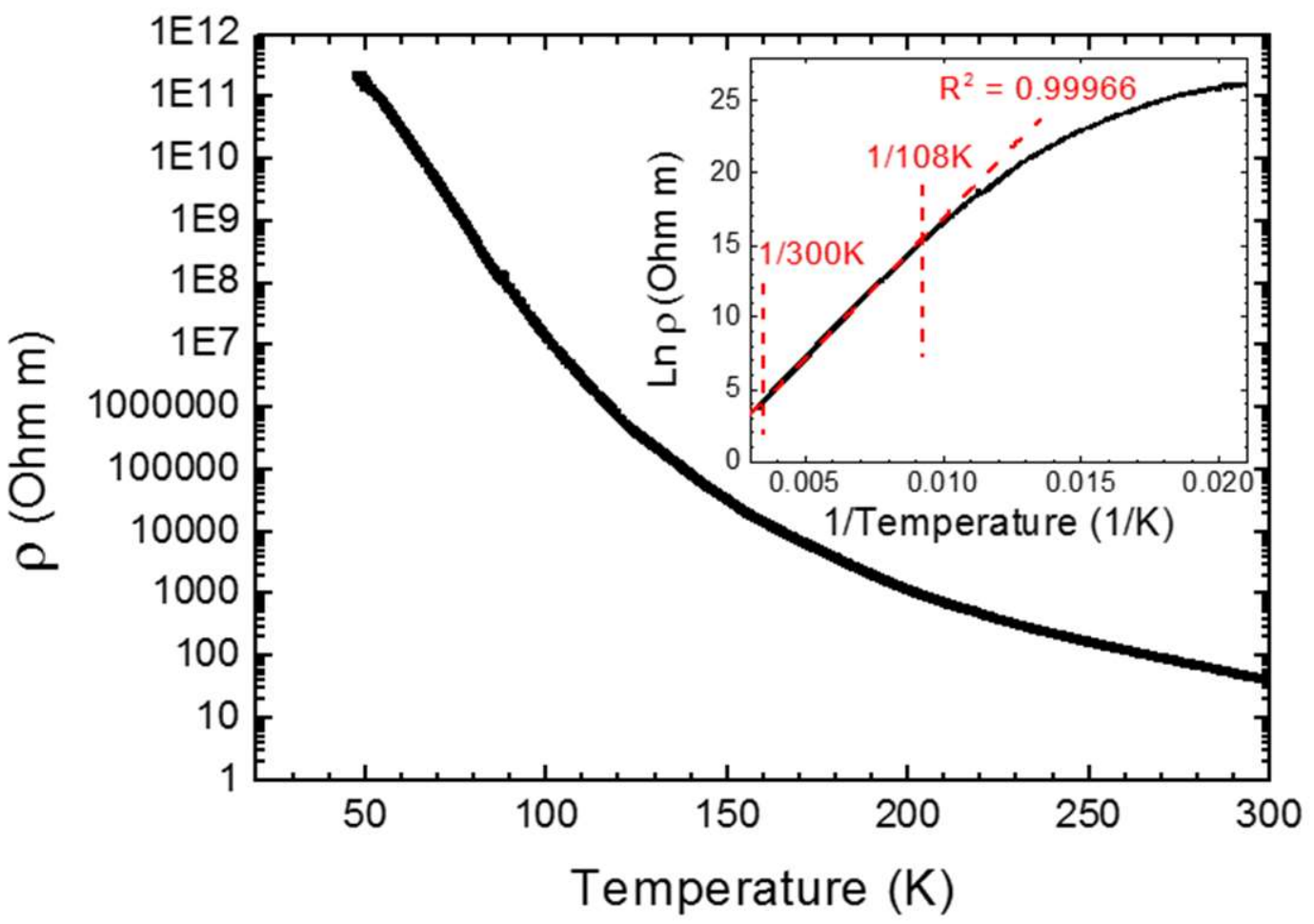

Figure 7 : Electrical resistivity $\rho$(T) of YMnO. Inset: semilogarithmic ln($\rho$) vs 1/T plot showing the thermally activated behavior predicted by the Arrhenius law (blue line). Red dashed lines indicate the T range of the fit.

*Capacitance and magnetization*

In the lack of a direct measurement of the ferroelectricity hysteresis cycle, it is necessary to combine pyroelectric measurements with capacitance (C) vs. temperature, since a ferroelectric materials must present an anomaly in the dielectric constant at the ferroelectric transition. The capacitance of YMnO disks vs. T at various frequencies is reported in figure 8. The obtained C values are not indicative due to the presence of a Schottky barrier in the junctions between the metallic electrodes and semiconductor [32] however we can comment on the qualitative behavior. The curves exhibit two small anomalies, at $T^*$ = 72 K and $T^*_{bis}$= 45 K. This supports the previous observation of a pyrocurrent peak centered at v with a shoulder at $T^*_{bis}$. The position of these anomalies is independent on the frequency while the intensity of the peak at $T^*$ increases with frequency and vanishes at the lowest frequency measured (1 kHz).

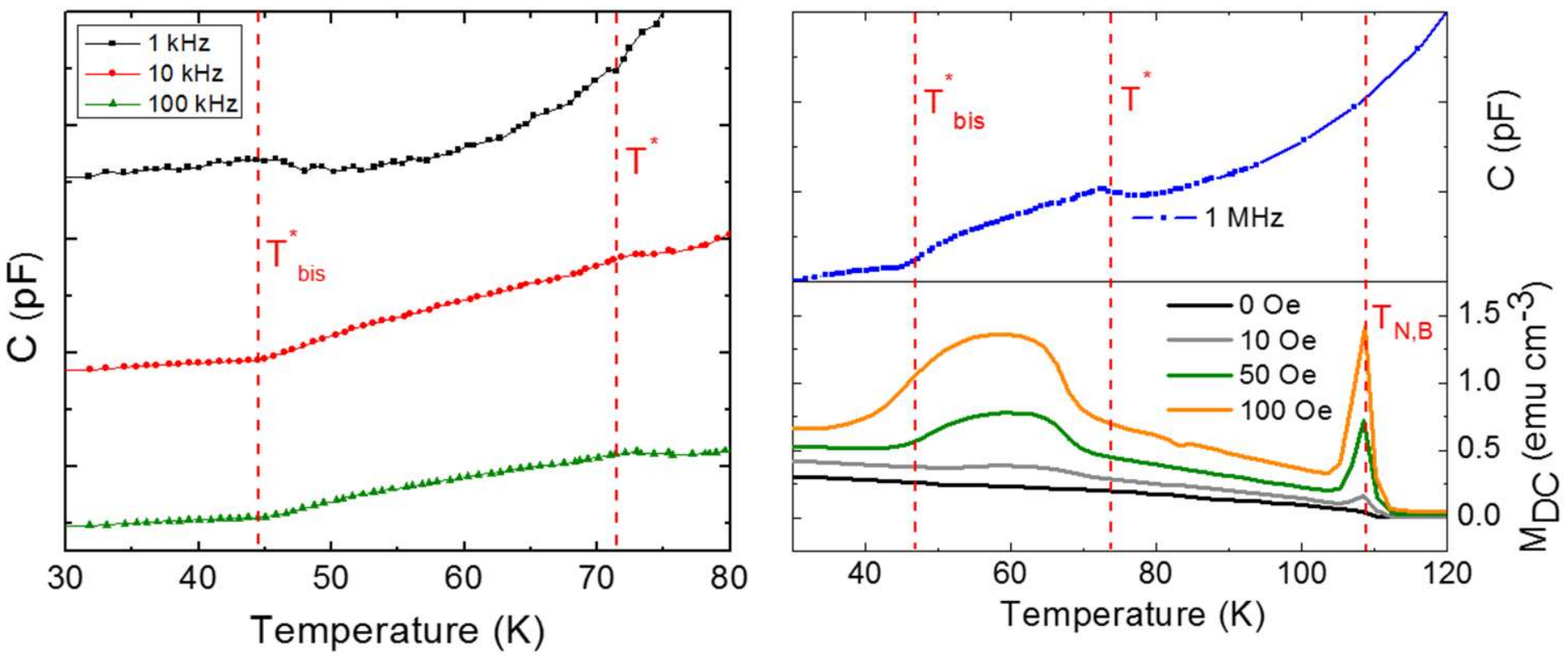

Figure 8 : Capacitance vs Temperature plot from impedance measurements performed with AC voltage of 200 mV with frequencies 1 kHz, 10 kHz, 100 kHz and 1 MHz. On the bottom of the right panel we have reported the DC M(T) curve of YMnO in order to make clear the concordance of the several critical temperature in both measurements.

We also have reported in the right-low panel of figure 8 the DC magnetic behavior of YMnO, as also reported in our previous work [15]. Remarkably, the two anomalies of C are located at the beginning and the end of the field-induced anomaly reported in M(T) between $T^*$ and $T^*_{bis}$. On the other hand, at the long-range magnetic order $T_{N,B}$, there is no anomaly in C for all frequencies. These results suggest that YMnO is an unusual improper ferroelectric where ferroelectricity is induced by the short-range magnetic phenomenon occurring between 72 and 45 K. In order to investigate the nature of this uncommon magnetic behavior we further performed AC-magnetization measurements as a function of frequency. The curve are reported in figure 9 and we clearly see a sharp cusp located around 66K slowly shifting towards higher temperature when increasing frequency. This behavior is similar to the one expected for a spin-glass behavior around its freezing temperature [33] Spin-glass behavior is known to be due to competition between FM and AFM interaction and has recently been reported in quadruple perovskite of manganese [34].

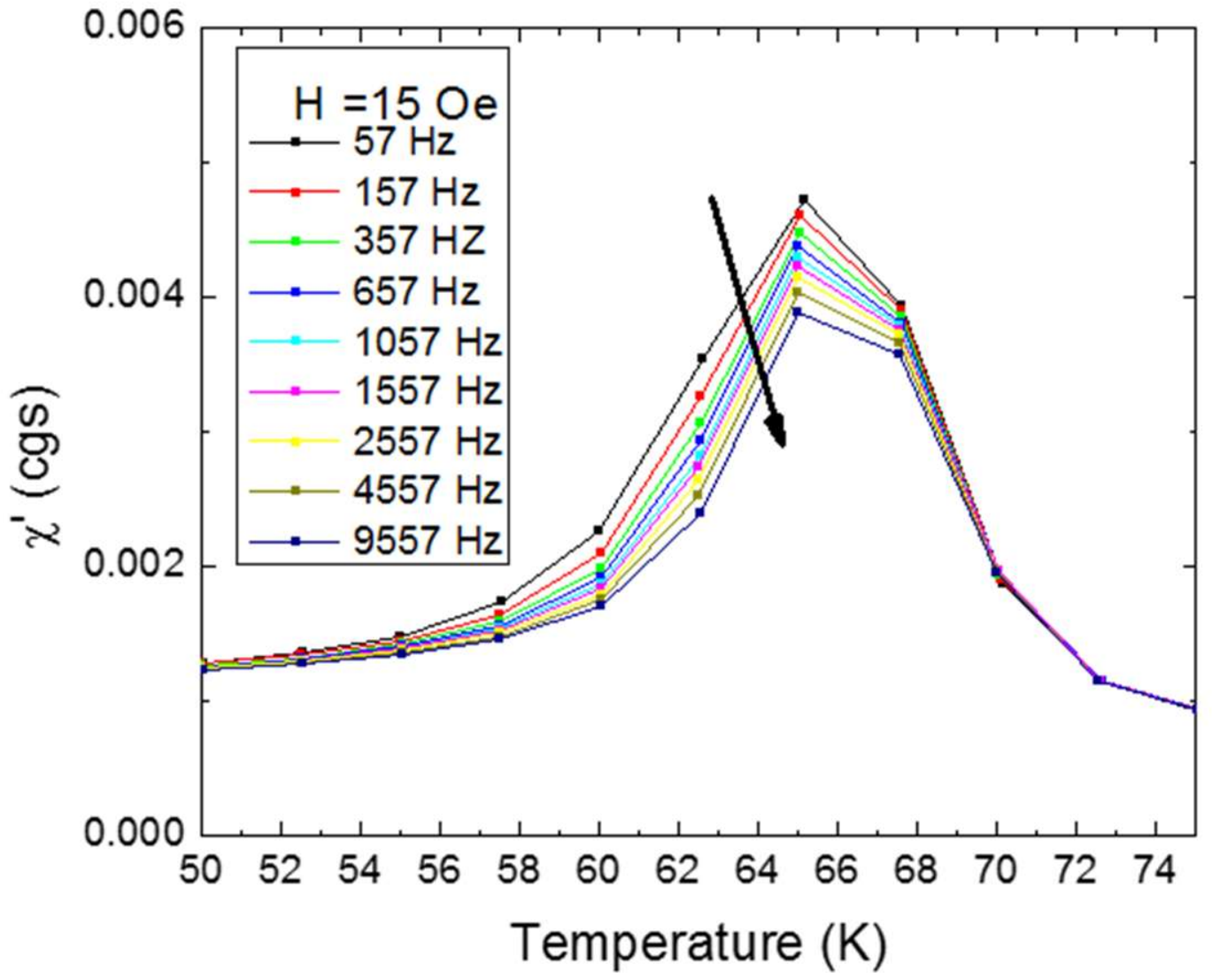

Figure 9 : Real part of the AC susceptibility of YMnO measured at different frequencies from 57 to 9557 Hz as indicated in the legend.

Recently, some authors have suggested other hypotheses to explain the magnetic properties of YMnO [17]. Their paper states that the first transition at 108K (named $T_{N,B}$ in our paper) is due to the ferrimagnetism of A'-sites which would orient simultaneously to the B Mn cations, while the second peak on the magnetic susceptibility should not be considered as an intrinsic response of YMnO but, on the contrary, considered as a spurious signal assignable to the impurity $YMn_3O_6$ [35]. As already discussed previously, none of the neutron diffraction data reported so far reaches the sufficient resolution needed to choose between the two models of magnetic structure, we maintain our simpler magnetic model of canted B-sites ordering at 108 K with no simultaneous ordering of A'-sites. In this case, the weak ferromagnetism behavior below $T_{N,B}$ of the system can be ascribed to a small difference of 0.004 μB in the magnetic moment of 4e and 4f B sites which is allowed by symmetry. Even though, from powder synchrotron diffraction reported as Supporting Information, we actually found 3% $YMn_30_6$ spurious phase in our sample powder, on the basis of our present AC magnetic data the hypothesis of an extrinsic magnetic response due to this impurity at 66K cannot be confirmed. Indeed, by assuming that the whole contribution to this magnetic peak is uniquely related to such impurity, we obtained a maximum value of the susceptibility of 232 $cm^3$/mol, much larger than the published value (about 5 $cm^3$/mol), measured in a pure $YMn_3O_6$ phase [35]. Hence, we conclude that this magnetic signal has to be considered a proper feature of YMnO. The low temperature broad magnetic peak detected by DC magnetization (inset of figure 8) suggests a dynamical process coupled with the formation of the electric polarization as it has the same critical temperatures. We hypothetically attribute the phenomenon to a short-range spin-glass-like ordering of the Mn ions on the A' sites of the perovskite going with the charge localization on Mn ion onto the A' site responsible for the net electric polarization below $T^* = 72$ K .

CONCLUSION

In conclusion, we have clarify the nature of the structural transition at $T_S = 200$ K in the small A-site quadruple perovskite $YMn_3Mn_4O_{12}$ (YMnO). Our diffraction data highlight the setting of a commensurate superstructure at $T_S$ and by means of single-crystal X-ray diffraction on high quality crystal, we were able to reveal the breaking of centrosymmetry by refining the structure in the *Ia* space-group. This results is in accord with all data already reported for this compounds and solve the previous inconsistency about the Yttrium position. As the aforementioned space-group allows for ferroelectricity, we have investigated in detail the pyrocurrent, transport, dielectric and magnetic properties of polycrystalline sample of YMnO over a wide temperature range. All these measurements point toward a very unique behavior of YMnO when lowering temperature: i) the centrosymmetry is broken at high temperature ($T_S$ = 200K), ii) then the long-range magnetic order of B-sites occurs at $T_{N,B}$ = 108 K and the compound simultaneously enters in an insulating dielectric state, iii) finally, a remnant polarization is stabilized concomitantly with a magnetic anomaly at $T^*$ = 70 K. We propose the assumption that YMnO is a peculiar magnetic ferroelectrics in which the ferroelectricity is driven by a short-range magnetic order. To the best of our knowledge, this is the first time that such a kind of behavior is reported in magnetic ferroelectrics.

## ASSOCIATED CONTENT

The Supporting Information is available free of charge on the ACS Publications website at DOI: 10.1021/acs.inorgchem.6b00774.
Details of XRPD, DSC, specific heat, magnetic, and dielectric measurements and impedance analysis (PDF)

(Word Style "TE_Supporting_Information"). **Supporting Information**. A listing of the contents of each file supplied as Supporting Information should be included. For instructions on what should be included in the Supporting Information as well as how to prepare this material for publications, refer to the journal's Instructions for Authors.

The following files are available free of charge.

brief description (file type, i.e., PDF)

brief description (file type, i.e., PDF)

## AUTHOR INFORMATION

### Corresponding Author

* marine.verseils@synchrotron-soleil.fr

### Present Addresses

† Synchrotron SOLEIL, 91190 Gif-sur-Yvette Cedex, France

**Author Contributions**

The manuscript was written through contributions of all authors. All authors have given approval to the final version of the manuscript.

**Funding Sources**

PhD “Leonardo Da Vinci” funding of the Franco-Italian University.

ACKNOWLEDGMENT

M.V. gratefully acknowledges financial support from the “Leonardo da Vinci” doctoral program of the Franco-Italian University and from synchrotron SOLEIL. The authors are grateful to P. Fertey for his assistance in single-crystal data integration.

REFERENCES

(Word Style “TF_References_Section”). References are placed at the end of the manuscript. Authors are responsible for the accuracy and completeness of all references. Examples of the recommended format for the various reference types can be found at http://pubs.acs.org/page/4authors/index.html. Detailed information on reference style can be found in *The ACS Style Guide,* available from Oxford Press.

BRIEFS (Word Style "BH_Briefs"). If you are submitting your paper to a journal that requires a brief, provide a one-sentence synopsis for inclusion in the Table of Contents.

SYNOPSIS (Word Style "SN_Synopsis_TOC"). If you are submitting your paper to a journal that requires a synopsis, see the journal's Instructions for Authors for details.